\documentclass[aps,prd,preprint,nofootinbib]{revtex4}

\usepackage{amsmath,amssymb}
\usepackage{graphicx}
\usepackage{slashed}
\usepackage{longtable}
\usepackage{color}

\newcommand{\be}{\begin{equation}}
\newcommand{\ee}{\end{equation}}
\newcommand{\ba}{\begin{eqnarray}}
\newcommand{\ea}{\end{eqnarray}}
\newcommand{\ban}{\begin{eqnarray*}}
\newcommand{\ean}{\end{eqnarray*}}
\newcommand \nn {\nonumber}

\begin{document}

\title{Effective Coupling Constant of Plasmons}

\author{Margaret E. Carrington}
\affiliation{Department of Physics, Brandon University,
Brandon, Manitoba, Canada\\
and Winnipeg Institute for Theoretical Physics, Winnipeg, Manitoba, Canada}

\author{Stanis\l aw Mr\' owczy\' nski}
\affiliation{Institute of Physics, Jan Kochanowski University, Kielce, Poland \\
and National Centre for Nuclear Research, Warsaw, Poland}

\date{August 15, 2019}

\begin{abstract}

We study an ultrarelativistic QED plasma in thermal equilibrium. Plasmons -- photon collective excitations -- are postulated to correspond not to poles of the retarded photon propagator but to poles of the propagator multiplied by the fine structure constant. This product is an invariant of the renormalization group that is  independent of an arbitrarily chosen renormalization scale. In addition, our proposal is physically motivated since one needs to scatter a charged particle off a plasma system to probe its spectrum of collective excitations. We present a detailed calculation of the QED running coupling constant at finite temperature using the Keldysh-Schwinger representation of the real-time formalism. We discuss the issue of how to choose the renormalization scale and show that the temperature is a natural choice which prevents the breakdown of perturbation theory through the generation of potentially large logarithmic terms. Our method could be applied to anisotropic systems where the choice of the renormalization scale is less clear, and could have important consequences for the study of collective modes. 

\end{abstract}

\maketitle

\section{Introduction}

The spectrum of collective excitations is a fundamental characteristic of any plasma system as it controls thermodynamic and transport properties of the system. In weakly coupled QED or QCD plasmas, non-trivial plasmon dispersion relations can be obtained perturbatively at the one-loop level. Quantities like the plasma frequency and the screening mass depend linearly  on $\alpha = e^2/4\pi$ for QED or $\alpha = g^2/4\pi$ for QCD. However, in both QED and QCD the coupling constant depends on a characteristic energy or momentum scale which emerges from the renormalization procedure -- one says that the coupling `runs.' We need to know the relevant energy scale at which to define the coupling constant when studying collective excitations. If the system is in equilibrium, there are  two possible scales to choose from -- the momentum of the mode under consideration, or the temperature of the plasma. The situation is even  less clear when we deal with non-equilibrium plasmas, where spectra of collective modes are often very rich, see {\it e.g.} \cite{Carrington:2014bla}. It is therefore of interest to develop a method to study this issue for non-thermalized systems. 

This subject is extremely important in the context of quark-gluon plasmas which are studied experimentally using relativistic heavy-ion collisions. The dynamics of the plasma is governed by QCD which is asymptotically free. This means that the plasma becomes weakly interacting at sufficiently high momentum scales, and perturbative  methods are applicable. 

We are particularly interested in non-equilibrium anisotropic QCD plasmas, which exist during some early phase of the evolution of the system that is produced in a heavy-ion collision. Anisotropic plasmas produce unstable modes, as discussed at length in the review \cite{Mrowczynski:2016etf}. These unstable modes are damped through inter-parton collisions, but the damping effect is higher order in the coupling constant. In the perturbative regime, the damping effect is small and unstable modes can play an important role. In a strongly coupled plasma, unstable modes would not have a significant effect on the dynamics of the system. The physics of the early-stage plasma therefore crucially depends on whether or not the regime of asymptotic freedom is reached. 

We intend to study this problem systematically in the context of quark-gluon plasma, both in and out of equilibrium. In this paper we will start with the simpler case of a thermal QED plasma using the real-time Keldysh-Schwinger formalism \cite{Chou:1987,Carrington:2006xj}, which is applicable to both equilibrium and non-equilibrium systems. Since quark masses are usually neglected in QCD plasma, we consider here an ultrarelativistic QED plasma where electrons and positrons are treated as massless. We focus on photon collective modes, which are called plasmons. To obtain their dispersion relations, we study the resumed retarded propagator at the one-loop level. 

In equilibrium systems, collective modes are usually studied at the leading order of the Hard-Thermal-Loop (HTL) approximation \cite{Le-Bellac-2000}, which assumes that the system's temperature is much bigger than the frequency and wave vector of the collective mode. In these calculations, it is conventional to not consider the effect of the polarization of the vacuum, since this contribution is subleading within the HTL approximation. The effective coupling emerges beyond leading order, as a result of vacuum polarization effects. In this paper, we will study this issue in a more general context, with particular emphasis on the question of how the renormalization scale should be chosen in an anisotropic system.

To explain the problem, let us discuss briefly the example of the thermodynamic pressure. A perturbative calculation produces a series of terms that depend on the renormalized coupling $e(\mu)$ and logarithms of the form $\ln(\mu/T)$, where $\mu$ is the renormalization scale and $T$ is the temperature. If we choose $\mu=T$, we obtain a simpler expression for the pressure with the logarithms set to zero and the surviving factors of $e(\mu)$ replaced by $e(T)$. It is commonly said that we resum to all orders the logarithm terms into a running coupling constant which is defined at the appropriate scale. This has been verified to order $e^5(\mu)$ by explicit calculation \cite{Frenkel:1992az,Arnold:1994eb,Braaten:1995jr}. We also note that if the renormalization scale $\mu$ were chosen to be some number much bigger or smaller than the temperature, the perturbative expansion would break down, due to the appearance of large log terms. For this reason one says that the choice $\mu=T$ prevents the appearance of large logs. The method to absorb the logarithm terms in the running coupling constant is discussed in the context of vacuum field theory in Chapter 18 of Ref.~\cite{Weinberg-II-1995}. We emphasize that in the calculation of a thermodynamic quantity, the only scale in the problem is the temperature and therefore it is natural to set the renormalization scale equal to the temperature, as discussed above. The study of plasmons is more complicated because more than one momentum scale comes into play. 

Physically meaningful quantities should be scale independent, but the renormalized photon propagator is not renormalization group invariant. However, the product of $e^2(\mu)$ and the propagator is. We therefore postulate that  plasmons correspond not to poles of the retarded photon propagator but to poles of this product. We stress that this proposal is also motivated physically, since one needs to scatter a charged particle off the plasma system to probe its spectrum of collective photon excitations. The product of the squared coupling and the propagator has the same structure as the pressure in terms of its dependence on the renormalization scale: it is given perturbatively by a series of terms depending on $e(\mu)$ and $\ln(\mu/T)$ that can be rewritten by absorbing all log terms into a coupling $e(T)$. 

Throughout the paper we use natural units where $\hbar = c = k_B = 1$. The indices $i,j,k = 1, 2, 3$ and $\mu, \nu = 0, 1, 2, 3$ label, respectively, the Cartesian spatial coordinates and those of Minkowski space. The signature of the metric tensor is $(+,-,-,-)$.

\section{Resummed photon propagator}
\label{sec-ret-prop}

\subsection{Vacuum propagator}
\label{vac-prop}

The resumed time-ordered propagator is defined through the Dyson-Schwinger equation as
\ba
\label{Dyson-Schwinger-eq}
\label{dyson}
\big(D^{-1}\big)^{\mu \nu}(k) 
= \big(D^{-1}_0\big)^{\mu \nu} (k) - \Pi^{\mu \nu}(k) ,
\ea
where $\Pi^{\mu \nu}(k)$ is the time-ordered self-energy or polarization tensor and $D^{\mu\nu}_0(k)$ is the free  propagator. We consider two gauges: the general covariant gauge (GCG) and temporal axial gauge (TAG). In these two gauges the non-interacting propagators are 
\ba
\label{D0-GCG}
\text{GCG:~~~~}  D_0^{\mu \nu} (k)  &=& 
\frac{1}{k^2 + i  0^+} \Big(g^{\mu \nu} - (1 - \xi) \frac{k^\mu k^\nu}{k^2} \Big) ,
\\[2mm]
\label{D0-TAG}
\text{TAG:~~~~} D_0^{\mu\nu}(k) &=& 
\frac{1}{k^2 + i  0^+} \Big(g^{\mu \nu} + (1 + \xi) \frac{k^\mu k^\nu}{(k \cdot n)^2} 
- \frac{k^\mu n^\nu + n^\mu k^\nu}{(k \cdot n)}\Big)  ,
\ea 
where $\xi$ is an arbitrary gauge parameter that physical results should be independent of and the four-vector $n^\mu = (1,0,0,0)$ defines the reference frame where the gauge condition is imposed. 

In vacuum, gauge and Lorentz invariance dictate that the self-energy depends only on $k^2$ (not $k$) and that it can be written as the product of a four-dimensionally transverse tensor and a scalar function $P(k^2)$ in the form
\be
\label{vac-pi}
\Pi_{\rm vac}^{\mu\nu}(k) = \left(g^{\mu\nu}k^2 - k^\mu k^\nu\right) P(k^2) .
\ee
In GCG inverting the Dyson-Schwinger equation (\ref{Dyson-Schwinger-eq}) gives
\be
\label{D-GCG-T=0}
D^{\mu\nu}(k) = \frac{1}{k^2\big(1- P(k^2)\big)} \left(g^{\mu\nu}-\frac{k^\mu k^\nu}{k^2}\right) 
+ \xi \frac{k^\mu k^\nu}{k^2} ,
\ee
and in strict  ($\xi=0$) TAG we obtain
\ba
\label{D-TAG-T=0}
- D^{ij}(k) = \frac{1}{k^2 \big(1-  P(k^2)\big)} \, T^{ij}({\bf k}) 
+ \frac{1}{k_0^2\big(1- P(k^2)\big)} \, L^{ij}({\bf k})  ,
\ea
where we have defined
\ba
\label{def-TL}
L^{ij}({\bf k}) \equiv \frac{k^i k^j}{{\bf k}^2}  ,
~~~~~~~~~~~~
T^{ij}({\bf k}) \equiv \delta^{ij} - \frac{k^i k^j}{{\bf k}^2}  \,.
\ea
From now on when we refer to TAG we always mean with the choice of the gauge parameter $\xi=0$. One advantage of working in TAG is that the components of the propagator that have time-like indices are identically zero, and the components with only spatial indices are decomposed in terms of two projection operators, one three-dimensionally transverse, $T({\bf k})$, and the other three-dimensionally longitudinal, $L({\bf k})$. 

The one-loop vacuum contribution to the self-energy can be calculated in Euclidean space using dimensional regularization. The procedure is standard and the result can be found in many textbooks, see {\it e.g.} Chapter 11.2 of Ref.~\cite{Weinberg-I-1995}. Rotating back to Minkowski space gives
\ba
P(k^2) &=& -\frac{e^2}{2\pi^2}\left[\frac{1}{6}\left(\frac{1}{\delta}-\gamma\right) + \int^1_0 dx\,x(1-x) 
\ln \left(\frac{4\pi M^2}{m_e^2-x(1-x)k^2}\right)\right] ,
\label{pre-dimreg}
\ea
where $M$ is a mass parameter introduced by the regularization procedure. The parameter $\delta$ is related to the dimension $d$ of the momentum space through $d=4-2\delta$, and $\gamma \approx 0.5772$ is the Euler-Mascheroni constant. In the formula (\ref{pre-dimreg}) we keep a finite electron mass $m_e$, as is usually done in vacuum QED. We note that when $k^2<0$ the argument of the logarithm is positive definite and $P(k^2)$ is real and analytic. When $k^2>0$ the logarithm has a branch cut when its argument becomes negative. The maximum value of the factor $x(1-x)$ for $x\in[0,1]$ is 1/4 and therefore the branch cut begins at $k^2=4m_e^2$, which is the threshold for particle-antiparticle production. The sign of the imaginary part is determined by including the appropriate infinitesimal imaginary regulator.

\subsection{Medium propagator}

Thermal field theory can be formulated in a Lorentz covariant way \cite{Weldon:1982aq}, nevertheless there is a preferred reference frame in the problem which is the rest frame of the heat bath, which we define with the four-vector $n^\nu = (1,0,0,0)$, as in Eq.~(\ref{D0-TAG}).  The reason TAG is particularly useful at finite temperature is that the gauge condition is imposed in the rest frame of the heat bath. 

An arbitrary symmetric tensor can no longer be decomposed in terms of the coefficients of two tensors, and we must extend the basis to include four independent tensors. Using the notation of Ref.~\cite{Le-Bellac-2000}, see section Sec.~5.2.2, we introduce the four-vector 
\be
n_T^\mu \equiv \left(g^{\mu\nu}-\frac{k^\mu k^\nu}{k^2}\right) n_\nu ,
\ee
which is the component of $n^\nu$ transverse to $k^\mu$, and define
\be
\label{4tensors}
\begin{split}
A^{\mu\nu}(k) &\equiv g^{\mu\nu}-\frac{k^\mu k^\nu}{k^2} - \frac{n_T^\mu n_T^\nu}{n_T^2} ,
~~~~~~~~~ 
B^{\mu\nu}(k) \equiv \frac{n_T^\mu n_T^\nu}{n_T^2} ,
\\[2mm] 
C^{\mu\nu}(k) &\equiv  k^\mu n_T^\nu+n_T^\mu k^\nu ,
~~~~~~~~~~~~~~~~~~~
E^{\mu\nu}(k) \equiv  \frac{k^\mu k^\nu}{k^2} .
\end{split}
\ee
The sum of $A$ and $B$ is the four-dimensionally transverse tensor
\be
\label{A+B}
A^{\mu\nu}(k) + B^{\mu\nu}(k) =  g^{\mu\nu}-\frac{k^\mu k^\nu}{k^2}.
\ee 
The tensors $A$ and $B$ are individually four-dimensionally transverse, and $A$ is three-dimensionally transverse. We note for future reference the relations
\be
T^{ij}({\bf k}) = - A^{ij}(k) ,
~~~~~~~~~
L^{ij}({\bf k})  = - \frac{k^2}{k_0^2} \, B^{ij}(k) 
= - \frac{k^2}{2k_0{\bf k}^2} \, C^{ij}(k) 
= \frac{k^2}{{\bf k}^2} \, E^{ij}(k)  .
\ee
The full multiplication table for the tensors (\ref{4tensors}) is shown in Table \ref{tab-mult}.
\begin{table}[b]
\begin{center}
\begin{tabular}{|c|c|c|c|c|}
\hline  
~~~~~~~~~~~~~& ~~~~~$A$ ~~~~~&~~~~~ $E$ ~~~~~&~~~~~ $B$~~~~~ &~~~~~ $C$~~~~~ \\ 
\hline
$A$ & $A$ & 0 & 0 & 0 \\ 
$E$ & 0 & $E$  & 0 & $C$\\ 
$B$ & 0 & 0  & $B$ & $C$ \\ 
$C$ & 0 & $C$  & $C$ &~~ -${\bf k}^2(B+E)$~~\\
\hline 
\end{tabular}
\end{center}
\caption{Multiplication table for the tensors defined in equation (\ref{4tensors}). \label{tab-mult}}
\end{table}

The polarization tensor is four-dimensionally transverse (due to the Ward identity) and therefore $\Pi^{\mu\nu}(k)$ can be decomposed using only the tensors $A$ and $B$ as
\be
\label{Pi-med}
\Pi^{\mu\nu}(k) = \Pi^T (k) \, A^{\mu\nu}(k) + \Pi^L (k) \, B^{\mu\nu}(k)  ,
\ee
where the indices $T$ and $L$ indicate the scalar coefficients of the tensors that are three-dimensionally transverse and longitudinal, respectively. 

We will calculate the retarded polarization tensor. We note that the Dyson-Schwinger equation (\ref{dyson}) gives the resummed retarded propagator in terms of the non-interacting retarded propagator and retarded polarization tensor, without coupling to other causal structures (which is not true if one works with the time-ordered propagator). Decomposing the free propagator in terms of tensors (\ref{4tensors}) and inverting the Dyson-Schwinger equation (\ref{Dyson-Schwinger-eq}) gives the propagator in GCG 
\ba
\label{D-CG}
D^{\mu\nu}(k) &=& D^T(k) \, A^{\mu\nu}(k) 
+  D^L(k) \, B^{\mu\nu}(k) +  \frac{\xi}{k^2 + i k_0 0^+} \, E^{\mu\nu}(k) ,
\ea
and in TAG
\ba
\label{D-TAG}
- D^{ij}(k) &=& D^T(k) \, T^{ij}({\bf k}) 
+ \frac{k^2}{k_0^2} \, D^L(k) \, L^{ij}({\bf k}) ,
\ea
where 
\be
\label{D-TL-def}
D^{T,L}(k) \equiv \frac{1}{k^2 - \Pi^{T,L}(k)} .
\ee
The positions of poles of the photon propagator should be independent of the chosen gauge and the formulas (\ref{D-CG}), (\ref{D-TAG}), and (\ref{D-TL-def}) are clearly consistent with gauge independence. In either gauge, a transverse plasmon is found as a solution of the dispersion equation $k^2 - \Pi^T(k) =0$ and a longitudinal plasmon is the solution of $k^2 - \Pi^L(k) =0$. 

In the next section we will show that the retarded polarization tensor can be divided into a vacuum piece, and a medium contribution that vanishes in the limit that the distribution function goes to zero. 
We write 
\ba
\label{Pi-vac-med}
\Pi^{T,L} (k) = \Pi_{\rm vac}^{T,L} (k) + \Pi_{\rm med}^{T,L} (k) ,
\ea
with 
\ba
\label{vac-lim}
\Pi_{\rm vac}^T (k) = \Pi_{\rm vac}^L (k) = k^2 P(k^2) .
\ea 

We comment briefly on the fact that we have discussed the time-ordered vacuum polarization tensor in Sec.~\ref{vac-prop}, while in this section we work with the retarded polarization tensor which is directly relevant to the study of collective modes. The time-ordered vacuum polarization tensor can be Wick rotated to Euclidean space where we can perform the dimensional regularization. On the other hand, the Dyson-Schwinger equation defined on the Keldysh-Schwinger contour can be most easily solved for the retarded polarization tensor, as explained above Eq.~(\ref{D-CG}). In Sec.~\ref{sec-renorm} we will need to combine the vacuum and medium contributions to the self-energy, but this is straightforward because the real parts of the time-ordered and retarded self-energies are equal to each other \cite{Le-Bellac-2000}, and the imaginary parts are finite and play no role in our calculation. In Eqs. (\ref{Pi-vac-med}) and (\ref{vac-lim}), and all following equations, it should be understood that $\Pi_{\rm vac}^{T,L}(k)$ will be assigned retarded boundary conditions.  

Using Eqs.~(\ref{Pi-vac-med}) and (\ref{vac-lim}), we can rewrite the transverse and longitudinal contributions to the propagator (\ref{D-TL-def}) as
\ba
\label{D-TL-def-2}
D^{T,L}(k) \equiv \frac{1}{k^2\big(1-P(k^2)\big) - \Pi_{\rm med}^{T,L}(k)} \,.
\ea
In the vacuum limit $\Pi_{\rm med}^{T,L}(k)\to 0$ and using equations (\ref{A+B}) and (\ref{Pi-med}) we recover the usual form for the vacuum polarization tensor (\ref{vac-pi}). We see also that equations (\ref{D-CG}) and (\ref{D-TAG}) reduce to (\ref{D-GCG-T=0}) and (\ref{D-TAG-T=0}). In addition, we have from equations (\ref{D-TAG}) and (\ref{D-TL-def-2}) that in vacuum the coefficient of the three-dimensionally transverse tensor has a pole when $k^2=0$, but the coefficient of the three-dimensionally longitudinal tensor has a pole at $k_0^2=0$.  Physically this tells us that in vacuum there is no propagating three-dimensionally longitudinal mode, and the appearance of these modes is a medium effect.

\section{Retarded polarization tensor}
\label{sec-ret-Pi}

Weldon computed \cite{Weldon:1982aq} the real part of the time-ordered polarization tensor of an ultrarelativistic QED plasma, which equals the real part of the corresponding retarded polarization tensor. He obtained the leading HTL contribution and the next-to-leading order contribution from the one-loop diagram. We have calculated the real part of the one-loop retarded polarization tensor, to next-to-leading order. We apply the Keldysh representation of the real-time formulation of statistical field theory. Our method does not explicitly require the use of thermal distribution functions, and we expect it to be generalizable to anisotropic systems that can be described by a distribution function which has the same asymptotic form as a thermal distribution. We have shown that in equilibrium our method reproduces the result of Ref. \cite{Weldon:1982aq}. Our calculation is described in Appendix \ref{retarded-all}, and further details can be found in Refs.~\cite{Carrington:2006xj,Carrington:2007zz}. A related calculation was done recently using an on-shell effective field theory approach \cite{Manuel:2016wqs,Carignano:2017ovz}.

The retarded polarization tensor is
\ba
\label{integrand-full}
\Pi^{\mu\nu}(k) = 2 e^2 \sum_{n=\pm 1}\int \frac{d^3p}{(2\pi)^3} \,
\frac{1-2n_f(|{\bf p}|)}{|{\bf p}|} \,
\frac{2p^\mu p^\nu + p^\mu k^\nu + k^\mu p^\nu - g^{\mu\nu}p\cdot k}{(p+k)^2 + i(p_0+k_0)0^+} 
\bigg|_{p_0=n |{\bf p}|} ,
\ea
where $n_f(|{\bf p}|) \equiv (e^{|{\bf p}|/T} +1)^{-1}$ is the distribution function of massless fermions. 

The vacuum and medium contributions can be easily separated in the expression (\ref{integrand-full}) as required by Eq.~(\ref{Pi-vac-med}). The real part of the vacuum piece is ultraviolet divergent and must be renormalized. In section \ref{vac-prop} we have renormalized the time-ordered vacuum self-energy, by performing a Wick rotation to Euclidean space and using dimensional regularization (as already mentioned, the real parts of the retarded and time-ordered self-energies are equal). The medium part is ultraviolet finite due to the distribution function which goes exponentially to zero as $|{\bf p}|$ approaches infinity. 

The polarization tensor (\ref{integrand-full}) is symmetric ($\Pi^{\mu\nu}(k) = \Pi^{\nu\mu}(k)$) and four-dimensionally transverse ($k_\mu \Pi^{\mu\nu}(k) = 0$). According to Eq.~(\ref{Pi-med}), such a tensor depends on only two independent scalar functions, which we have called $\Pi^T(k)$ and  $\Pi^L(k)$ and defined in Eq.~(\ref{Pi-med}). The easiest way to obtain $\Pi^T(k)$ and  $\Pi^L(k)$ is to calculate the zero-zero component and the trace of the polarization tensor, and to use the relations
\be
\label{conversion}
\Pi^T(k) = \frac{1}{2}\left(\Pi^\mu_{~\mu}(k) +\frac{k^2}{{\bf k}^2}\Pi^{00}(k) \right) ,
~~~~~~~
\Pi^L(k)  = -\frac{k^2}{{\bf k}^2}\Pi^{00}(k) .
\ee

We have calculated the medium part of the one-loop retarded polarization tensor to next-to-leading order in the expansion in $(k_0/T,|{\bf k}|/T)$. We give some details of our method in Appendix \ref{nlo-fin}. To leading order one obtains the familiar HTL results 
\be
\label{medium-lo}
\begin{split}
\big[\Pi^{00}_{\rm med}(k)\big]_{\rm LO} &= \frac{e^2T^2}{3} 
\left( 1-\frac{k_0}{2 |{\bf k}|} \ln
\left|\frac{|{\bf k}|+k_0}{|{\bf k}|-k_0}\right|  - i \pi \Theta(-k^2)\right)  ,
\\[2mm]
\big[\Pi^{\mu}_{{\rm med} \, \mu}(k)\big]_{\rm LO} &= \frac{e^2T^2}{3} .
\end{split}
\ee
The next-to-leading order contributions are
\be
\label{medium-nlo}
\begin{split}
& \big[\Pi^{00}_{\rm med}(k)\big]_{\rm NLO} 
= \frac{e^2 {\bf k}^2}{12 \pi ^2} \ln \left(\frac{\sqrt{8} k^2}{T^2}\right) ,
\\[2mm]
& \big[ \Pi^\mu_{{\rm med} \, \mu}(k) \big]_{\rm NLO} 
= -\frac{e^2 k^2}{4 \pi ^2} \ln \left(\frac{4 k^2}{T^2}\right) .
\end{split}
\ee
We note that equations ({\ref{medium-lo}}) and (\ref{medium-nlo}) show that there is a non-zero imaginary part when $k^2<0$. This is exactly the opposite behavior from what was found in vacuum (see equation (\ref{pre-dimreg})) where we saw that the imaginary part is non-zero only for time-like momenta, in which case the virtual photon can decay into physical final states. For the medium contribution, the non-zero imaginary part of the self-energy appears for space-like momenta and corresponds physically to the scattering of electrons and positrons with momenta of order $T$ on the low-momentum photon. In plasma physics, this phenomenon is known as Landau damping. We also note that the results (\ref{medium-lo}) and (\ref{medium-nlo}) agree to next-to-leading order with those given in \cite{Weldon:1982aq}.

\section{Renormalization}
\label{sec-renorm}

Renormalization is usually done by including counterterms in the Lagrangian and  re-expressing bare quantities in terms renormalized ones. The divergent part of the counter-terms are chosen to cancel the divergences in the $n$-point functions, and the finite parts are determined by enforcing a chosen renormalization condition. We write the renormalized propagator 
\ba
\label{Z3-def}
\hat{D}^{T,L}(k,\mu) \equiv \frac{1}{Z_3(\mu)}D^{T,L}(k) 
= \frac{1}{k^2\big(1 - \hat{P}(k^2,\mu) \big) - \Pi^{T,L}_{\rm med}(k)} ,
\ea
where the first part of the equation defines the renormalization constant $Z_3$, and $\mu$ is a new scale that enters through the renormalization condition. We distinguish renormalized quantities from their non-renormalized counterparts by adding hats to the former ones and explicitly showing their dependence on $\mu$. Our renormalization condition is that in the vacuum limit, where the medium part of the polarization tensor vanishes, the renormalized propagator with $k^2 \rightarrow -\mu^2$ coincides with the free propagator. This condition can be enforced by shifting
\ba
\label{shifted}
P(k^2) \to \hat{P}(k^2,\mu) = P(k^2) - P(-\mu^2) ,
\ea
which gives $\hat{\Pi}^{\mu\nu}_{\rm vac}(k)\big|_{k^2=-\mu^2} = 0$ and provides the renormalization constant as
\be
\label{Z3-mu}
Z_3(\mu) = 1 + P(-\mu^2) .
\ee 

We note that to obtain the results in Sec.~\ref{sec-ret-Pi} for the medium contribution to the self-energy we set the electron mass to zero, because we are interested in an ultrarelativistic plasma where the electron mass is assumed negligible compared to the temperature. In the vacuum calculation that we are discussing here, we must therefore also set $m_e=0$ for consistency. The $m_e\to 0$ limit is conventional in the vacuum calculation anyway, because the integrals that are calculated are dominated by their ultraviolet contributions. With $m_e=0$ equations (\ref{pre-dimreg}) and (\ref{shifted}) give
\ba
\label{shifted2}
\hat{P}(k^2,\mu) = \frac{e^2}{12\pi^2}\ln \left(\frac{-k^2}{\mu^2}\right) .
\ea
We mention again that the argument of the logarithm in Eq.~(\ref{shifted2}) indicates that the self-energy has an imaginary part for $k^2>0$, which means physically that a virtual time-like photon can decay into physical final states. Using retarded boundary conditions we take $k^2 \to k^2+i k_0 0^+$ and rewrite the formula (\ref{shifted2}) as
\ba
\label{shifted3}
\hat{P}(k^2,\mu) = \frac{e^2}{12\pi^2}
\left[\ln \left(\frac{|k^2|}{\mu^2}\right) - i\pi\Theta(k^2) \,{\rm sgn}(k_0) \right].
\ea

Substituting the expression (\ref{shifted3}) into Eq.~(\ref{Z3-def}) provides
\ba
\label{ren-DT-0}
\hat{D}^{T,L}(k,\mu) &=&
\frac{1}{k^2 \big[1 - \frac{ \hat{e}^2(\mu)}{12\pi^2} \big(
\ln\big(\frac{|k^2|}{\mu^2}  \big) - i\pi\Theta(k^2)\,{\rm sgn}(k_0) \big)\big] - \Pi^{T,L}_{\rm med}(k)} ,
\ea
where the medium contributions are given in equations (\ref{medium-lo}) and (\ref{medium-nlo}). The couping constants in (\ref{shifted3}) are renormalized coupling constants (which are defined in the next section) and should be properly written as a function of the scale $\mu$. This is made explicit in equation (\ref{ren-DT-0}), and the coupling constants in the medium contribution in this equation should also be taken to be renormalized coupling constants. Using equations (\ref{medium-lo}) and (\ref{medium-nlo}) and replacing $\hat{e}^2(\mu)$ by $4\pi \hat{\alpha}(\mu)$ we obtain
\be
\label{ren-DT}
\hat{D}^{T,L}(k,\mu) = 
\frac{1}{k^2 \big[1 - \frac{\hat{\alpha}(\mu)}{3\pi} 
\ln\big(\frac{T^2}{\mu^2}\big) \big] - \hat{\alpha}(\mu) \, \pi^{T,L}(k)} ,
\ee
where $\pi^{T,L}(k)$ indicates contributions without potentially large logarithmic factors, which are therefore of no interest to us. We stress that $\pi^{T,L}(k)$ are defined so that they do not include the overall factor $\hat\alpha(\mu)$, and should not be confused with $\Pi^{T,L}(k)$.

\section{Renormalization group invariant and collective modes}

We postulate that collective modes correspond not to poles of the photon propagator but to poles of the photon propagator multiplied by the fine structure constant $\alpha$. One motivation is that physical quantities should be independent of the  renormalization scale $\mu$, and the product $\hat{\alpha} (\mu) \, \hat{D}^{L,T}(k,\mu)$ is a renormalization group invariant, as discussed at length in Chapter 9 of the classical text \cite{Bogoliubov-Shirkov-1980}. Our idea is also motivated physically: to probe a photon collective mode one needs to scatter a charged particle off the plasma system. The charge should therefore be considered together with the propagator from which the collective mode will be determined. 

In vacuum it is natural to choose the renormalization scale to be equal the momentum scale $\sqrt{|k^2|}$, which is the only physical scale available. We will show below that at finite temperature one should choose $\mu=T$. 

We start by deriving the equation for the running coupling constant, which describes how the coupling constant evolves with the scale at which it is defined. To satisfy the Ward identity the charge must be renormalized using the introduced previously renormalization constant $Z_3(\mu)$ given by Eq.~(\ref{Z3-mu})
\be
\label{e-ren-def}
 \hat{\alpha}(\mu) = Z_3(\mu) \, \alpha . 
\ee
Since the bare coupling constant on the right side of  Eq.~(\ref{e-ren-def}) is independent of $\mu$, the evolution of $\hat{\alpha} (\mu)$ is determined by the equation 
\be
\label{evol-eq}
\mu \frac{d \hat{\alpha} (\mu)}{d\mu} = \beta (\mu),
\ee
where the beta function is defined as
\be
\label{beta-def}
\beta(\mu) \equiv 
\mu \frac{d Z_3(\mu)}{d\mu} \frac{\hat{\alpha} (\mu)}{Z_3(\mu)} .
\ee
Using Eq.~(\ref{Z3-mu}) together with the formula (\ref{pre-dimreg}), the well-known beta function at the one-loop level is found as
\be
\label{beta-1-loop}
\beta (\mu) = \frac{2}{3\pi} \, \hat{\alpha}^2 (\mu) ,
\ee 
and the solution of the evolution equation (\ref{evol-eq}) equals
\be
\label{alpha-evol}
\hat{\alpha} (\mu) =
\frac{\hat{\alpha} (\mu_0)}{1 - \frac{\alpha (\mu_0)}{3\pi} \ln\big(\frac{\mu^2}{\mu_0^2}\big) } ,
\ee
with the famous Landau pole structure. 

Equations (\ref{Z3-def}) and (\ref{e-ren-def}) clearly show that the product $\alpha (\mu) \,  D^{T,L}(k,\mu)$ is a renormalization group invariant. We see this explicitly by writing
\ba
\nn
 \hat{\alpha} (\mu) \,  \hat{D}^{T,L}(k,\mu) 
&=& \frac{\hat{\alpha} (\mu)}{1  -  \frac{\hat{\alpha}(\mu) }{3 \pi} \ln\big(\frac{T^2}{\mu^2}\big)} \,
\frac{1}{k^2 - \hat{\alpha}(\mu) \, \pi^{T,L} (k)  } 
\\ \label{prod-inv-1}
&=& \hat{\alpha}(T)\frac{1}{k^2 - \hat{\alpha}(\mu) \, \pi^{T,L} (k)} ,
\ea
where the first line is obtained using Eq. (\ref{ren-DT}) and the equality holds up to ${\cal O}(\hat{\alpha}^2)$. The factor $\hat{\alpha}(T)$ in the second line comes from Eq.~(\ref{alpha-evol}), which also makes it clear that the remaining factor $\hat{\alpha}(\mu)$ can be written as $\hat{\alpha}(T)$ since the difference between these two quantities is of order $\hat{\alpha}^2$, and we have already dropped terms of this order. Equation (\ref{prod-inv-1}) therefore becomes
\ba
\label{prod-inv-final-2}
 \hat{\alpha}(\mu) \,  \hat{D}^{T,L} (k,\mu) 
= \frac{\hat{\alpha}(T)}{k^2 - \hat{\alpha}(T) \, \pi^{T,L} (k)}
= \hat{\alpha} (T) \,  \hat{D}^{T,L}(k, T)  ,
\ea
which shows that the natural renormalization scale of collective modes in the thermal plasma is the temperature.

\section{Conclusions and Outlook}

We claim that dispersion relations of plasmons should not be calculated by finding the poles of the retarded photon propagator, but rather by finding poles of the propagator multiplied by the fine structure constant. This product is a renormalization group invariant, as the amplitude of a physical process should be. We have  given a physical argument to use this product to define plasmons. We have also shown that the statement that $\hat{\alpha}(\mu)\hat{D}^{T,L}(k,\mu)$ is a renormalization group invariant is equivalent to the statement that the polarization tensor is given perturbatively by a series of terms depending on $\hat{\alpha}(\mu)$ and $\ln\left(\frac{\mu}{T}\right)$ that can be rewritten by absorbing all log terms into a running coupling $\hat{\alpha}(T)$ using Eq.~(\ref{alpha-evol}). 

We are going to extend the analysis presented in this paper to the case of fermionic collective modes in QED plasma, which are conventionally determined by the poles of the retarded electron propagator. In this case, however, the propagator multiplied by the coupling constant is not a renormalization group invariant, but instead that which is independent of the renormalization scale is the product of electron propagator and the vertex function. 

Our ultimate goal is to study collective modes of QCD plasma, with a particular emphasis on anisotropic systems. The structure of anisotropic plasmas is much more complicated and the choice of the renormalization scale is not clear, but our method is rather general and hopefully can be applied to such a system.

\section*{Acknowledgments}

We are very grateful to Alina Czajka and Erik Kofoed for numerous fruitful discussions. This work was partially supported by the National Science Centre, Poland under grant 2018/29/B/ST2/00646, and by the Natural Sciences and Engineering Research Council of Canada.

\appendix

\section{Retarded polarization tensor}
\label{retarded-all}

In this Appendix we give some details of our calculation of the one-loop retarded polarization tensor (\ref{integrand-full}). 
We work in the real-time formulation of finite temperature field theory, using the Keldysh representation. The one-loop contribution to the retarded polarization tensor can be found rather easily starting with what is often called the 1-2 basis, see {\it e.g.} Sec.~IVA of \cite{Czajka:2010zh}, but we sketch here a more general method reviewed in \cite{Carrington:2006xj} which is applicable to multi-loop diagrams.

The electron propagator is a 2$\times$2 matrix of the form
\ba
G & = & \left(\begin{array}{cc}
G_{rr} &G_{ra} \\
G_{ar} & G_{aa} \\
\end{array}\right)
= \left(\begin{array}{cc}
G_{\rm sym} &G_{\rm ret} \\
G_{\rm adv} & 0 \\
\end{array}\right) ,
\ea
where the retarded, advanced and symmetric propagators are given by
\ba
\nn
G_{\rm ret}(p)  &=& (\slashed{p} + m_e)r(p),
\\ 
\label{prop-def1}
G_{\rm adv}(p) &=& (\slashed{p} + m_e) a(p),
\\ \nn
G_{\rm sym}(p) &=& (\slashed{p} + m_e) f(p),
\ea
with
\ba
\nn
r(p)  &\equiv& \frac{1}{p^2-m_e^2+i0^+ {\rm sgn}(p_0)} ,
\\ [2mm]
\label{prop-def2}
a(p) &\equiv& \frac{1}{p^2-m_e^2-i0^+ {\rm sgn}(p_0)} ,
\\ [2mm] \nn
 f(p) &\equiv& -2\pi i 
\big[ \big(1-2 n_f({\bf p}) \big) \Theta(p_0) +  \big(1-2 \bar n_f(-{\bf p}) \big) \Theta(-p_0) \big]
\delta(p^2 - m_e^2) ,
\ea
where $n_f({\bf p})$ and  $\bar n_f({\bf p})$ are the fermion and anti-fermion distribution functions. The self-energy has the form
\ba
\Pi &=& \left(\begin{array}{cc}
\Pi_{rr} & \Pi_{ra} \\
\Pi_{ar} & \Pi_{aa} 
\end{array}\right)
= \left(\begin{array}{cc}
0 & \Pi_{\rm adv} \\
\Pi_{\rm ret} & \Pi_{\rm sym} 
\end{array}\right)
\ea
and the vertex function is a 2$\times$2$\times$2 tensor which can be written 
\ba
\Gamma^\mu &=& -ie \gamma^\mu \left(
\begin{array}{cc}
 \{\Gamma_{rrr},\Gamma_{rra}\} & \{\Gamma_{rar},\Gamma_{raa}\} \\
 \{\Gamma_{arr},\Gamma_{ara}\} & \{\Gamma_{aar},\Gamma_{aaa}\} \\
\end{array}
\right) = -ie \gamma^\mu \left(
\begin{array}{cc}
 \{0,1\} & \{1,0\} \\
 \{1,0\} & \{0,1\} \\
\end{array}
\right) .
\ea

The contribution to the retarded self-energy from the one-loop diagram is
\ba
i \Pi_{\rm ret}^{\mu \nu} (k) = i \Pi_{ar}^{\mu \nu}(k) = \frac{(-ie)^2}{2}\sum_{ii'jj'}\int \frac{d^4p}{(2\pi)^4} \,
\gamma^\mu \Gamma_{aij} \, G_{ii'}(p+k)G_{jj'}(p) \, \gamma^\nu \Gamma_{ri'j'} . 
\ea
The sum over Keldysh indices $\{i,i',j,j'\}\in\{r,a\}$ is easily done because $G_{aa}=0$ and a vertex function with an odd number of $a$ indices vanishes.  The result is 
\ba
\label{A5} 
\Pi_{\rm ret}^{\mu\nu} (k) 
= i\frac{e^2}{2}
\int \frac{d^4p}{(2\pi)^4} \,
{\rm Tr}\big[\gamma^\mu (\slashed{p} + m_e) \gamma^\nu (\slashed{k}+\slashed{p} + m_e)\big] 
\big[f(p) r(p+k)+a(p) f(p+k)\big] .
\ea
From this point on we assume that the system is in thermal equilibrium, we set the electron mass to zero, and we take $n_f({\bf p}) = \bar n_f(-{\bf p}) =  (e^{|{\bf p}|/T} +1)^{-1}$. Performing the trace over gamma matrices, and changing variables to combine the two terms in the square bracket in Eq.~(\ref{A5}), one obtains
\ba
\label{A6}
\Pi_{\rm ret}^{\mu\nu}(k) = 4ie^2\int \frac{d^4p}{(2\pi)^4}
\big[2p^\mu p^\nu + p^\mu k^\nu + k^\mu p^\nu - g^{\mu\nu}p\cdot(p+k)\big]f(p)r(p+k) .
\ea
It is straightforward to perform the integral over $p_0$ using the delta function in the symmetric propagator (see Eq.~(\ref{prop-def2})). The resulting expression for the photon self-energy is given in Eq. (\ref{integrand-full}).

\section{medium contribution}
\label{nlo-fin}

We present here some details of our calculation of  the medium contribution to the retarded polarization tensor in Eq.~(\ref{integrand-full}). The integrals of interest are
\ba
\Pi_{\rm med}^{00}(k) &=& -\frac{2e^2}{\pi^2}
\int_0^\infty d|{\bf p}|\,|{\bf p}|\,n_f(|{\bf p}|)\;\sum_{n=\pm 1} \frac{1}{2}\int_{-1}^1 dx \;I^{00}  ,
\\[2mm]
\Pi^{\mu}_{{\rm med}\,\mu}(k) &=& -\frac{2e^2}{\pi^2}
\int_0^\infty d|{\bf p}|\,|{\bf p}|\,n_f(|{\bf p}|)\;
\sum_{n=\pm 1} \frac{1}{2}\int_{-1}^1 dx \;I^{\mu}_{~\mu} ,
\ea
with
\ba
\nonumber 
I^{00}  &\equiv&  \frac{n |{\bf p}| k_0+2 {\bf p}^2  + {\bf p}\cdot {\bf k}}
{2 \left(n |{\bf p}| k_0 - {\bf p}\cdot {\bf k}\right)  + k^2 + i n 0^+} ,
\\ [2mm] \nonumber 
I^{\mu}_{~\mu} &\equiv& -\frac{2 \left(n |{\bf p}| k_0 - {\bf p}\cdot {\bf k}\right)}{2 \left(n |{\bf p}| k_0 - {\bf p}\cdot {\bf k} \right) + k^2+ i n 0^+} ,
\ea
where $x \equiv \frac{{\bf p}\cdot {\bf k} }{|{\bf p}| |{\bf k}|}$. 
To calculate $\Pi_{\rm med}^{00}(k) $ and $\Pi^{\mu}_{{\rm med}\,\mu}(k)$ to leading order we expand $I^{00}$ and $I^{\mu}_{~\mu}$ in $(k_0/|{\bf p}|,|{\bf k}|/|{\bf p}|)$ which gives 
\ba
I^{00}_{\rm LO} &=&  -\frac{2 {\bf p}^2 k^2}{\left(2 n |{\bf p}| k_0 -2 {\bf p}\cdot{\bf k} +i n 0^+\right)^2} 
+\frac{2 {\bf p}^2 + n |{\bf p}| k_0+{\bf p}\cdot{\bf k}}{2 n |{\bf p}| k_0  -2 {\bf p}\cdot{\bf k}+ i n 0^+} , \nonumber
\\ [2mm] \nn
I^\mu_{~\mu \,{\rm LO}} &=&  \frac{2 |{\bf k}| x-2 n k_0}{2 n k_0 -2 |{\bf k}| x +i n 0^+} .
\ea
Performing the $x$ integral gives the familiar HTL results (\ref{medium-lo}). 

As a check of our notation we observe that equation (\ref{conversion}) shows that 
\ba
\lim_{k_0\to 0} \frac{1}{k^2-\Pi^L_{\rm med}(k)} 
= \lim_{k_0\to 0} \frac{{\bf k}^2}{k^2({\bf k}^2+\Pi^{00}_{\rm med}(k))}
\ea
and from (\ref{medium-lo}) we find the pole at imaginary $|{\bf k}|=i m_D$ that corresponds to the screening mass $m_D^2 = e^2T^2/3$.

We find the NLO contribution from
\ba
\label{nlo1}
\Pi_{\rm NLO}(k) = -\frac{2e^2}{\pi^2}\int_0^\infty d|{\bf p}|\, |{\bf p}|\, n_f(|{\bf p}|)
\Big[\sum_{n=\pm 1} \frac{1}{2}\int_{-1}^1 dx \,(I -I_{\rm LO})\Big] .
\ea
The square bracket in equation (\ref{nlo1}) will be denoted $\chi^{00}$ or tr$\chi$. There are two kinds of terms in the integrand, those of the  form $(A-B x + i0^+)^{-1}$, and those with the form  $(A-B x+i0^+)^{-2}$. We separate the contributions from these two types of terms by writing $\chi^{00} = \chi^{00}_1+\chi^{00}_2$ and ${\rm tr}\chi = {\rm tr}\chi_1+{\rm tr}\chi_2$. It is straightforward to show that $\chi_2^{00}=1$ and ${\rm tr}\chi_2=0$.  Performing the $x$ integral for the type 1 terms, we obtain
\ba
\chi^{00}_1 &=& \frac{1}{2 |{\bf p}| \left(\omega _--\omega _+\right)}
\Big[
\left(|{\bf p}|-\omega _-\right) \left(|{\bf p}|-\omega _+\right) \left(\ln \left(|{\bf p}|-\omega _- + i 0^+ \right)
-\ln \left(|{\bf p}|-\omega _++i 0^+ \right)\right) \nonumber 
\\\label{B6a}
&&  ~~~~
- \left(|{\bf p}|+\omega _-\right) \left(|{\bf p}|+\omega _+\right) \left(\ln \left(|{\bf p}|+\omega _-+i 0^+ \right)
-\ln \left(|{\bf p}|+\omega _++i 0^+ \right)\right) \Big] , 
\\[2mm]
{\rm tr}\chi_1 &=& \frac{\omega_- \omega_+}{|{\bf p}| \left(\omega _--\omega _+\right)}\Big[
\ln \left(|{\bf p}|-\omega _-+i 0^+ \right)
- \ln \left(|{\bf p}|+\omega _- + i 0^+  \right)
\nonumber \\ 
\label{B6b}
 && ~~~~~~~~~~~~~~~~~ 
-\ln \left(|{\bf p}|-\omega _++i 0^+ \right)  
+  \ln \left(|{\bf p}|+\omega _+ +i  0^+ \right)\Big] ,
\ea
where we have defined $\omega_\pm \equiv (k_0\pm |{\bf k}|)/2$. We rewrite the arguments of the logs using the relations 
\ba
&& |{\bf p}| - \omega _\pm +i 0^+ 
=|{\bf p}| - \omega _\mp + i 0^+ \mp |{\bf k}| ,  \nonumber 
\\
&&   |{\bf p}| + \omega _\pm +i 0^+  
= |{\bf p}| + \omega_\mp +i 0^+ \pm |{\bf k}| , \nonumber
\ea 
and expanding the logarithms in $|{\bf k}|/(|{\bf p}|\pm \omega_\pm + i 0^+)$, Eqs.~(\ref{B6a}) and (\ref{B6b}) become  
\ba
\nn
\chi^{00}_1 &=& -1 
- \frac{\omega _- \left(\omega _+-\omega _-\right)}{4 \left({\bf p}^2-\omega _-^2+i 0^+ \right)}
+\frac{\omega _+ \left(\omega _+-\omega _-\right)}{4 \left({\bf p}^2-\omega _+^2+i 0^+ \right)}
\\[2mm]
&& 
\label{chi-1-00}
- \frac{\left(\omega _--\omega _+\right)\left(2 \omega _-+\omega _+\right)}{12 \left({\bf p}^2-2 \omega _-^2+i 0^+
   \right)}+\frac{\left(\omega _--\omega _+\right) \left(\omega _-+2 \omega
   _+\right)}{12 \left({\bf p}^2-2 \omega _+^+i 0^+ \right)} , 
\\[4mm]
\label{chi-1-tr}
{\rm tr}\chi_1  &=& -\frac{\omega _- \omega _+}{{\bf p}^2-\omega _-^2+i 0^+ }-\frac{\omega _- \omega
   _+}{{\bf p}^2-\omega _+^2+i 0^+ } .
\ea   
The term $-1$ on the right side of the first line in equation (\ref{chi-1-00}) cancels the contribution from $\chi^{00}_2$. The remaining terms in (\ref{chi-1-00}) and (\ref{chi-1-tr}) are substituted into Eq.~(\ref{nlo1}).

The last step is to perform the remaining integral over $|{\bf p}|$. We use the identity $n_f(|{\bf p}|) = n_b(|{\bf p}|)-2n_b(2|{\bf p}|)$, where $n_f(|{\bf p}|)$ and $n_b(|{\bf p}|)$ are fermionic and bosonic equilibrium distribution functions, and rescale variables  so that all terms have a factor $n_b(|{\bf p}|)$.  Finally we take the real part of each self-energy component and use 
\ba
\int_0^\infty d|{\bf p}| \, |{\bf p}|\, n_B(|{\bf p}|) \, {\cal P}\frac{1}{{\bf p}^2-M^2} = -\frac{1}{4}\ln\left[\frac{T^2}{M^2}\right] + \cdots
\ea
where $M$ is assumed positive and real, and the dots indicate terms higher order in $M/T$. 
The final next-to-leading order result is given in Eq.~(\ref{medium-nlo}).


\end{document}